\documentclass[manuscript]{aastex}

\newcommand{\vsi}{$v \sin i~$}
\newcommand{\ha}{H${\alpha}~$}

\usepackage{epsfig}
\usepackage{graphicx}
\usepackage{natbib}
\usepackage{amsmath}
\shorttitle{$\Omega$-slow, High $k$ Solutions for Line-Driven Winds}
\shortauthors{Silaj et al.}

\begin{document}

\title{Line-Driven Winds Revisited in the Context of Be Stars: \\ 
$\Omega$-slow Solutions with High $k$ Values }

\author{J.\ Silaj\altaffilmark{1}, M.\ Cur\'{e}\altaffilmark{2}, \and
  C.\ E.\ Jones\altaffilmark{1}}

\altaffiltext{1}{Department of Physics and Astronomy, The University
  of Western Ontario, London, Ontario, N6A 3K7, Canada}
\altaffiltext{2}{Instituto de F\'{i}sica y Astronom\'{i}a, Facultad
  de Ciencias, Universidad de Valpara\'{i}so, Av. Gran Breta\~{n}a
  1111, Casilla 5030, Valpara\'{i}so, Chile}

\begin{abstract}
The standard, or fast, solutions of m-CAK line-driven wind theory
cannot account for slowly outflowing disks like the ones that surround
Be stars.  It has been previously shown that there exists another
family of solutions --- the $\Omega$-slow solutions --- that is
characterized by much slower terminal velocities and higher mass-loss
rates. We have solved the one-dimensional m-CAK hydrodynamical
equation of rotating radiation-driven winds for this latter solution,
starting from standard values of the line force parameters ($\alpha$,
$k$, and $\delta$), and then systematically varying the values of
$\alpha$ and $k$.  Terminal velocities and mass-loss rates that are in
good agreement with those found in Be stars are obtained from the
solutions with lower $\alpha$ and higher $k$ values.  Furthermore, the
equatorial densities of such solutions are comparable to those that
are typically assumed in ad hoc models.  For very high values of $k$,
we find that the wind solutions exhibit a new kind of behavior.

\end{abstract}


\section{Introduction}

Current line-driven wind theory is a modification of CAK-theory ---
the theory describing the mass loss due to radiation force in hot
stars as originally developed by \citet{cas75} --- and is therefore
known as m-CAK theory.  While the original theory characterized the line
force with just two parameters --- $k$, a constant related to the
effective number of lines contributing to the driving force, and
$\alpha$, an index defining the mix of optically thick and
optically thin lines --- several important additions and alterations
were later made that significantly improved the model.  \citet{abb82}
refined the theory by considering a greatly expanded line list, and
introduced a third parameter, $\delta$, which describes the change in
ionization throughout the wind. \citet{fri86} and \citet{pau86}
introduced the finite disk correction factor, representing the star as
a uniform bright disk instead of as a point source.  With these
improvements, the model has been very successful at predicting the
mass-loss rates ($\dot{M}$) and wind terminal velocities ($v_\infty$)
from very massive (i.e., O-type) stars.  Additionally, it has been
used to describe other phenomena such as planetary nebulae and
quasars.  Attempts to apply the theory of line-driven winds to Be
stars to explain the formation of their circumstellar disks, however,
have only met with limited success.  In fact, a full understanding of
the exact mechanism(s) responsible for removing material from the
central star has not yet been attained.  Instead, a common approach to
modeling Be star disks has been to assume an ad hoc density structure
that can reproduce the observables.  One scenario that has been
developed extensively is the case in which the density distribution in
the equatorial plane falls as an $R^{-n}$ power law, following the
models of \citet{wat86}, \citet{cot87}, and \citet{wat87}, who found
$n \approx 2.0-3.5$ by comparing to observations of the infrared
excesses of Be stars.

\subsection{Classical Be Stars}

Be stars, sometimes referred to as Classical Be stars to distinguish
them from other B-type stars with emission such as the Herbig Ae/Be
stars and B[e] stars, are near-main-sequence (i.e., luminosity class
III-V) B-type stars that appear essentially normal in terms of their
gravity, temperature, and composition, but possess a geometrically
thin, gaseous, circumstellar decretion disk about their equator.  The
presence of the disk produces emission lines in the optical to the
near IR region of the spectrum, including one or more members of the
Balmer series, He~\textsc{i} lines, and, typically, various metal
lines as well.  The \ha emission line tends to be the strongest of the
emission features, and it is often modeled to obtain the average disk
properties since it is formed over a large region in the disk.

Be stars also possess a polar wind --- this is inferred from their UV
lines, which are absorbed in a low hydrogen density medium with
expansion velocities of the order of $\sim$2000~km\,s$^{-1}$.  The
winds of normal B-type stars attain similar velocities, but while the
mass-loss rates of the earliest B-type stars are of the order of
10$^{-8}$ to 10$^{-9}$ $M_{\sun}$\,yr$^{-1}$, Be stars have mass-loss
rates of the order of 10$^{-7}$ to 10$^{-9}$ $M_{\sun}$\,yr$^{-1}$
\citep{und82}.

A fundamental aspect of Be stars is their very rapid rotation, though
they do not appear to be rotating at their critical rotation rates.
\citet{por96} found the distribution of rotation rates to peak at
values around 70\%--80\% of the critical rate. \citet{tow04} argued
that the effects of equatorial gravity darkening could mean that the
\vsi of Be stars was systematically underestimated, and that they may
actually be rotating much closer to their critical velocities than
this, but still found their rotation to be subcritical.  The rapid
rotation appears to be of fundamental importance in the formation of
the disk, but its exact role is, as yet, not entirely clear.

Be stars are variable objects, displaying both short-term and
long-term variations in the appearance of their spectral
lines. Significant line profile variability, such as dips or bumps
travelling through the profiles, on timescales from hours to several
days, can be explained by nonradial pulsation (NRP, \citealt{baa82}).
Long-term variations, such as phase changes from the Be phase to a
normal B-star phase (and back again) are usually on timescales of
several years to several decades, and can be readily interpreted as
the formation and destruction of disks.  Other long-term variations
can occur within the Be phase, such as transitions between singly and
doubly peaked H$\alpha$ line profiles, or even a transformation
between a Be and Be-shell phase.  Changes of this nature are most
commonly understood as structural changes in the disk.  Finally,
approximately one-third of Be stars exhibit $V/R$ variations, which
present as a cyclic asymmetry with changing peak heights in the violet
and red components of the emission lines. The one-armed disk
oscillation model of \citet{oka91, oka96} has proven successful in
replicating these variations, and is generally accepted as the current
best model to describe this aspect of Be stars.  To date, however,
there exists no physical model that fully describes all aspects
of the dynamic nature of Be star disks simultaneously.

\subsection{Overview of Previous Work}

Many attempts at employing line-driven wind theory to describe the Be
star phenomenon have been made.  Some early examples include
\citet{mar84}, who considered the effects of rotation in a
radiation-driven stellar wind and applied the results to Be stars,
\citet{poe86} who employed a rotating, magnetic, radiation-driven wind
model to Be stars, and \citet{dea89}, who included rotation effects
and the simulation of viscous forces in the equations of motion and
applied the resulting model to Be stars.  The wind-compressed disk
(WCD) model of Be stars by \citet{bjo93} --- a two-dimensional
hydrodynamic model that proposed a meridional current compressing and
confining the equatorial material into a very thin disk --- was
initially very promising, but later calculations by \citet{owo96} that
included the nonradial line force components and the effects of
gravity darkening showed that these two factors can inhibit the
formation of a WCD structure.

While m-CAK theory properly describes the polar wind of Be stars, a
recurring problem that pervades the models described above is that they
exhibit large equatorial expansions that result in terminal
velocities that are too large, typically of the order of
$\sim$1000~km\,s$^{-1}$. A hydrodynamical model by \citet{ste94} found
a peak separation of 2000~km\,s$^{-1}$ between the $V$ and $R$ peaks
of their model H$\alpha$ profile, representative of the excessive
width of H$\alpha$ profiles computed from radiative wind models.
According to \citet{poe78}, the fitting of H$\alpha$ profiles requires
terminal velocities of the order of $\sim$200~km\,s$^{-1}$.

\citet{cur04} found a new physical solution to the one-dimensional
nonlinear m-CAK hydrodynamic equation --- the $\Omega$-slow solution
--- that possesses a higher mass-loss rate and much lower terminal
velocity (roughly one-third of the standard solution's terminal
velocity).  Furthermore, this solution only emerges when the star's
rotational velocity is larger than $\sim$75\% of the critical
velocity.  This offers a natural explanation of Be stars: the poles
correspond to a nonrotational case, and therefore exhibit a fast
outflowing, low-density wind, while the fast rotation across the
equator causes the emergence of a wind characterized by a greater mass
outflow and lower terminal velocity.

In this work, we solve the one-dimensional (1D) hydrodynamic equation
for the $\Omega$-slow solution, and compare the resultant mass-loss
rates and terminal velocities with those typically assumed for Be
stars. In Section~2, we provide a brief recapitulation of the momentum
equation of the wind, as well as an overview of a typical ad hoc model
used to describe Be stars.  In Section~3, we numerically solve the
hydrodynamic equations for the $\Omega$-slow solutions.  We start from
the line-force parameters of \citet{abb82}, and then consider $\alpha$
and $k$ as free parameters in a systematic fashion, by first varying
$\alpha$ while holding the other line-force parameters constant, and
then repeating the exercise for various values of $k$. We show the
emergence of a new behavior for wind solutions with high $k$
values. Additionally, we compare the equatorial density structure
obtained from the line-driven wind solution with the ad hoc power-law
distribution that is the standard assumption for Be star disks, and we
show the resultant H$\alpha$ profiles (computed by using the two
different equatorial density structures as input for the code
\textsc{Bedisk}, and allowing the vertical structure to be computed by
the code under the assumption of approximate hydrostatic
equilibrium). In Section~4, we provide a summary, and discuss our
conclusions and future work.

\section{Theory}

The m-CAK model for line-driven winds requires that we solve both the
radial momentum equation,
\begin{equation}
\label{momentumeq}
v\frac{dv}{dr} = -\frac{1}{\rho}\frac{dp}{dr}
-\frac{GM(1-\Gamma)}{r^2} + \frac{v^2_{\phi}(r)}{r} +
g^{\rm{line}}(\rho,\frac{dv}{dr},n_E)
\end{equation}
and the equation of mass conservation,
\begin{equation}
\label{masseq}
\dot{M} = 4\,\pi\,r^2\,\rho\,v.
\end{equation}

In these expressions, $v$ is the fluid velocity, $\rho$ is the mass
density, $p$ is the fluid pressure, $\Gamma$ is Eddington parameter,
$v_{\phi} = v_{\rm{rot}}R_*/r$, where $v_{\rm{rot}}$ is the star's rotational
speed at the equator, and $g^{\rm{line}}(\rho,dv/dr,n_E)$ is the
acceleration due to the lines.  The line-force term given by
\citet{abb82}, \citet{fri86}, and \citet{pau86} is:
\begin{equation}
\label{lineforce}
g^{\rm{line}} =
\frac{C}{r^2}f_D\left(r,v,\frac{dv}{dr}\right)\left(r^2v\frac{dv}{dr}\right)^{\alpha}\left(\frac{n_E}{W(r)}\right)^{\delta}.
\end{equation}
The coefficient $C$ depends on $\dot{M}$, $W(r)$ is the dilution
factor, and $f_D$ is the finite-disk correction factor.  The reader is
referred to \citet{cur04} for a detailed derivation and the full
definitions of all variables, constants and functions. A key result of
Cur\'{e}'s work was that by introducing the coordinate change $u =
-R_*/r, w = v/a$, and $w' = dw/du$, with $a_{\rm{rot}} =
v_{\rm{rot}}/a$ where $a$ is the isothermal sound speed, the momentum
equation becomes
\begin{equation}
\label{momentumeq2}
F(u,w,w') \equiv \left(1 - \frac{1}{w^2}\right)w\frac{dw}{du} + A +
\frac{2}{u} + a^2_{\rm{rot}}u -
C'\,f_D\,g(u)(w)^{\delta}\left(w\frac{dw}{du}\right)^{\alpha} = 0.
\end{equation}

The standard method of solving this nonlinear differential equation
(Equation~\ref{momentumeq2}) and obtaining $C'(M)$ (the eigenvalue) is
to require the solution to pass through a critical point, defined as
the roots of the singularity condition,
\begin{equation}
\label{singularity}
\frac{\partial}{\partial w'}F(u,w,w') = 0,
\end{equation}
together with a constraint at the stellar surface (i.e., a lower
boundary condition) whereby the density is set to a specific value,
\begin{equation}
\label{density_bc}
\rho(R_*) = \rho_*.
\end{equation}
A regularity condition, namely,
\begin{equation}
\label{regularity}
\frac{d}{du}F(u,w,w') = \frac{\partial F}{\partial u} + \frac{\partial
  F}{\partial w}w' = 0,
\end{equation}
is also imposed at the critical point in order to find a physical wind
solution. As shown in \citet{cur04}, the analysis of these equations
revealed the existence of a new family of singular points, and proved
that the standard m-CAK solution vanishes at high rotational
velocities.

In this work, we use a high rotational velocity
($v_{\rm{rot}}/v_{\rm{crit}} = 0.90$) for our central star, and find
the solutions to the above equations (solutions that necessarily come
from the new family of solutions) with the
\textsc{Hydwind}\footnote{This code, which is described in detail in
  \citet{cur04}, will henceforth be referred to as \textsc{Hydwind}.}
code to obtain the equatorial density structure of the wind as a
function of radial distance from the central star.  This density
structure is compared with the ad hoc scenario whereby the equatorial
density is assumed to follow a simple power-law distribution,
\begin{equation}
\label{plmodel}
\rho(r,0) = \rho_0 \left(\frac{r}{R_*}\right)^{-n},
\end{equation}
where $\rho_0$ is the initial density of the disk at the stellar
surface and $n$ is the index of the (radial) power law.  Typical
values for these parameters are $\sim1 \times 10^{-12}$ g\,cm$^{-3}$
to $1 \times 10^{-10}$ g\,cm$^{-3}$ for $\rho_0$, and $\sim2.0-4.0$
for $n$ (see, e.g., \citealt{jon08}). \citet{sil10}, who modeled 56
H$\alpha$ spectra with the above described model, showed $\rho_0$
values of $5 \times 10^{-11}$ g\,cm$^{-3}$ and $1 \times 10^{-10}$
g\,cm$^{-3}$, and an $n$ value = 3.5, were strongly preferred by the
fits performed in that study.  Because a $\rho_0$ value of $1 \times
10^{-10}$ g\,cm$^{-3}$ is considered to be quite dense, we therefore
set $\rho_0$ in Equation (\ref{plmodel}) equal to a more mid-range value
of $5 \times~10^{-11}$ g\,cm$^{-3}$.  We also set $\rho_*$ in
Equation (\ref{density_bc}) equal to that same value to maintain as much
consistency as possible between the two models.  For completeness, we
have investigated the effects of increasing $\rho_*$ to $1 \times
10^{-10}$ g\,cm$^{-3}$, as well as decreasing it to $1 \times
10^{-11}$ g\,cm$^{-3}$, and we find that there is no impact on the
resulting mass-loss rates or terminal velocities of the solutions in
either case.

The equatorial density structure computed from each of the approaches
described above was supplied to the radiative transfer code
\textsc{Bedisk}. For a given equatorial density,
  \textsc{Bedisk} computes the vertical disk density under the
  assumption of isothermal hydrostatic equilibrium.  It then solves
  the equation of radiative transfer to obtain the disk temperature
  and level populations, iterating at each grid point to produce a
  self-consistent model of the physical conditions in the disk.
  \textsc{Bedisk} assumes that the disk is in pure Keplerian rotation;
  thus, our resultant models exhibit this velocity structure regardless
  of how the equatorial density was computed.  The aim of this work is
  to test if the material delivered to equatorial regions from our new
  wind solutions is sufficient to produce line emission.  An
  additional mechanism would be required to supply sufficient torque
  to allow the disk material to attain Keplerian velocities.

Once \textsc{Bedisk} has computed the full disk structure,
  line profiles are simulated by solving the transfer equation along
  lines of sight parallel to the star's rotation axis
  (i.e., essentially determining the amount of flux that would be
  received from the star if it were viewed pole-on), and then
  projecting that flux at different angles to simulate changing the
  observer's line of sight to the star. The interested reader is
referred to \citet{sig07} for a full description of \textsc{Bedisk}.

\section{Results}

We adopt the same parameters for a B1V star as in \citet{cur04}:
$T_{\rm{eff}} = 25\,000$~K, log~$g$ = 4.03, and $R = 5.3\,R_\sun$.
Similarly to the aforementioned work, we begin our analysis from the
line-force parameters of \citet{abb82}: $k = 0.3, \alpha = 0.5$, and
$\delta = 0.07$.  The rotational speed ($\Omega =
v_{\rm{rot}}/v_{\rm{crit}}$) of the central star is set to 0.90.

Table~\ref{table:B1a} shows the mass-loss rates and terminal
velocities for various $\alpha$ values when the other line-force
parameters are held constant.  The first line of the table corresponds
to the line-force parameters given in \citet{abb82}.  Starting from
this solution, we systematically lower the value of $\alpha$, which
represents physically a greater contribution of optically thin lines.
The table clearly shows that small changes in $\alpha$ have a large
impact on the mass-loss rate and terminal velocity, with smaller
$\alpha$ equating directly to smaller mass-loss rates and lower
terminal velocities. Interestingly, we could find no solutions for
$\alpha < 0.30$.

\begin{figure}
\plottwo{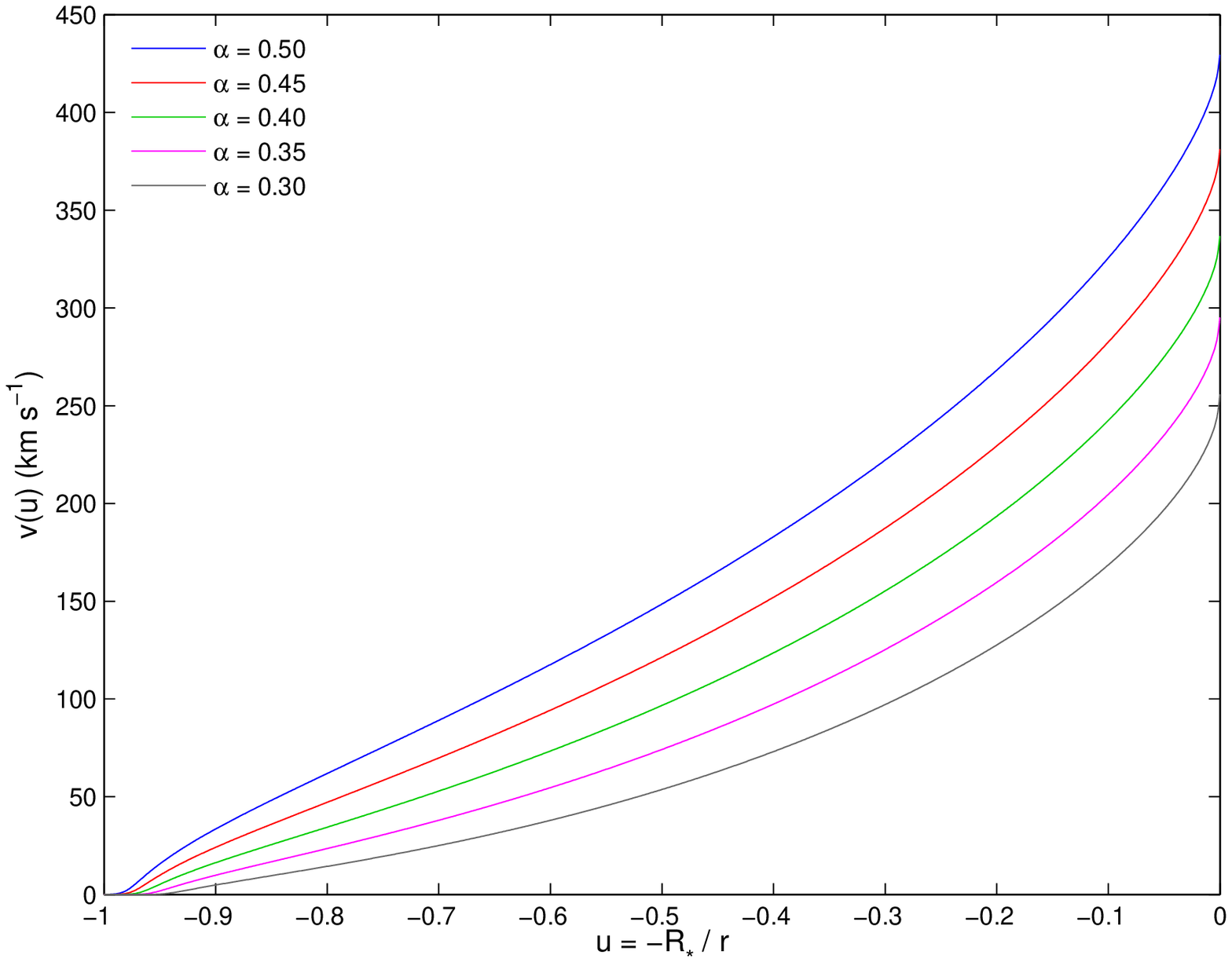}{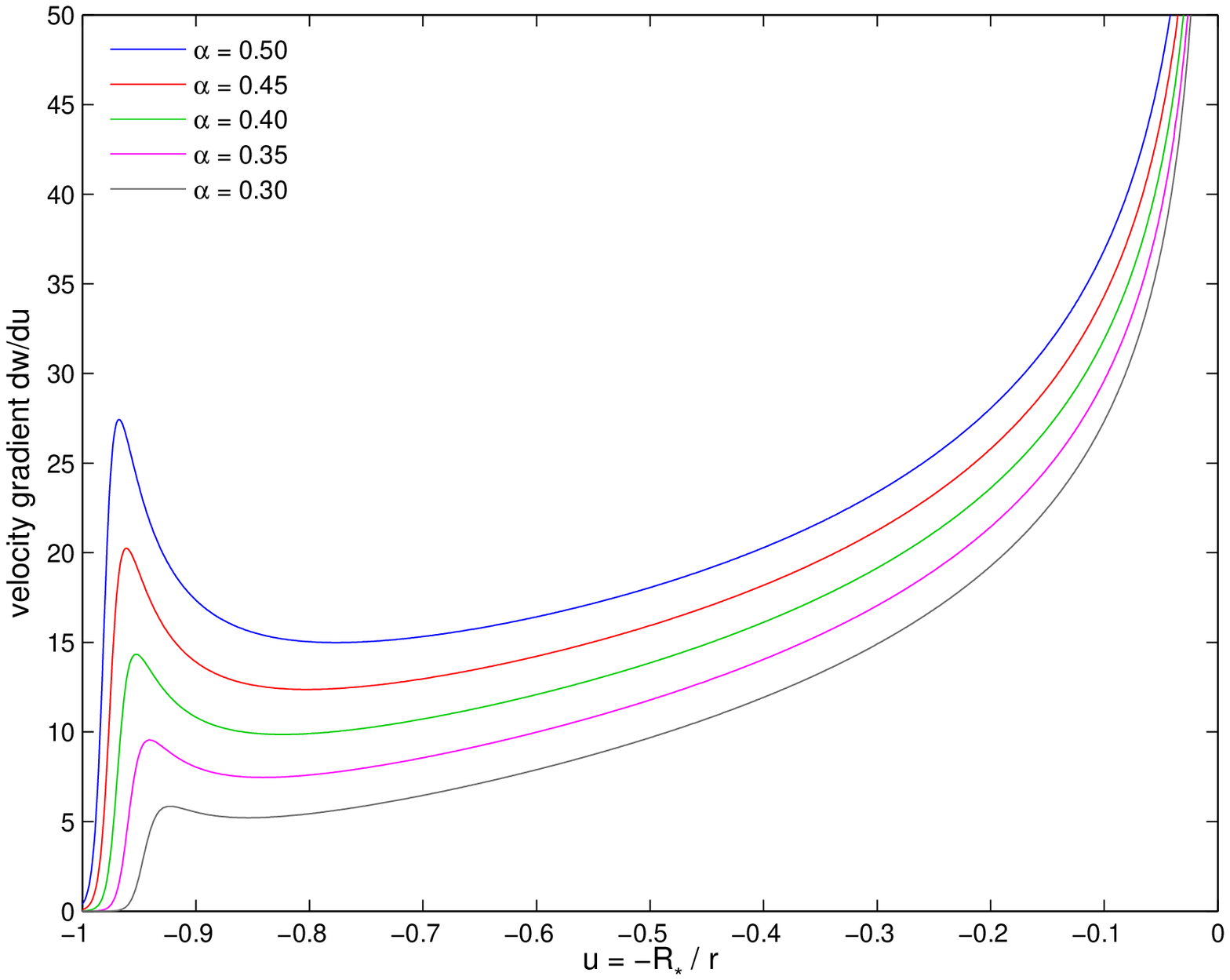}
\caption{Left panel: velocity profiles as a function of the inverse
  radial coordinate $u$, for fixed $k$ = 0.3 and $\delta$ = 0.07
  values, and various $\alpha$ values as indicated in the legend.
  Right panel: velocity gradients ($dw$/$du$) as a function of $u$ for
  the same solutions presented in the left panel.
\label{fig:alpha}}
\end{figure}

Figure~\ref{fig:alpha} illustrates the effect of lowering $\alpha$
from its starting value of 0.5 to 0.3, in steps of 0.05, on both the
velocity structure (left panel) and its gradient (right panel) as a
function of the inverse radial coordinate $u$.  Clearly, lowering
$\alpha$ corresponds to lower terminal velocities, but the overall,
characteristic shape of the velocity profile is preserved for all
values of $\alpha$ that we employed.  Similarly, the velocity gradient
shows less change in velocity (a smaller ``hump'') occurring close to
the stellar surface ($u = -1$) for lower $\alpha$ values, but retains
its basic shape overall.

Table~\ref{table:B1b} is similar to Table~\ref{table:B1a}, but with
$\alpha$ and $\delta$ now held constant while $k$ is varied. Again,
the first line in the table corresponds to the line-force parameters
of \citet{abb82}, and thus it is identical to the first line in
Table~\ref{table:B1a}.  From these results, it is immediately clear
that varying $k$ has only a marginal impact on the terminal
velocity.  Indeed, the only wind characteristic that $k$
significantly affects is the mass-loss rate, with greater $k$ values
corresponding to greater mass-loss rates.  Physically, a larger $k$
value represents a greater number of lines effectively contributing to
the driving of the wind.  When $k$ values greater than 9.0 were
employed, no solutions could be found.

\begin{figure}
\plottwo{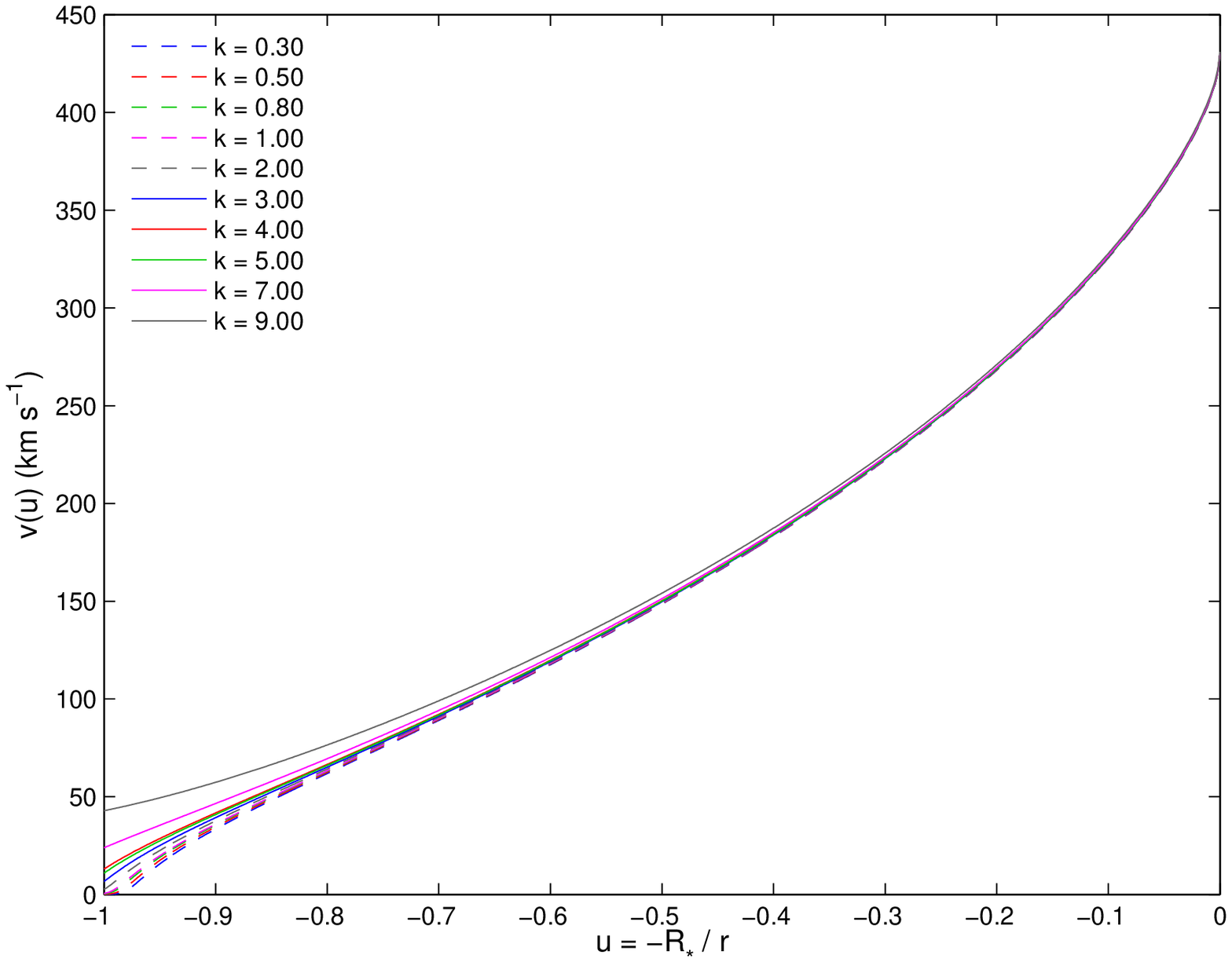}{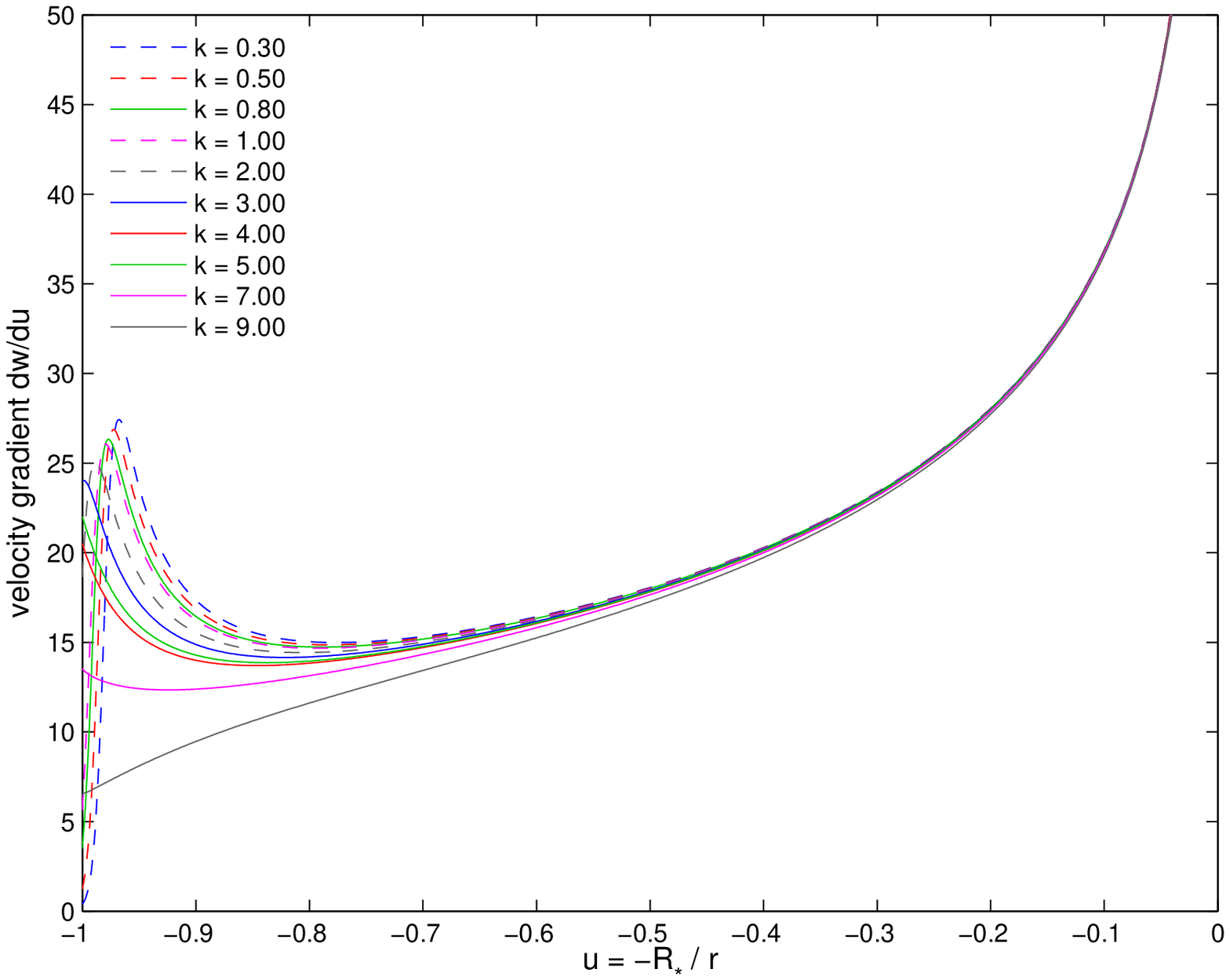}
\caption{Left panel: velocity profiles as a function of the inverse
  radial coordinate $u$, for fixed $\alpha$ = 0.5 and $\delta$ = 0.07
  values, and various $k$ values as indicated in the legend.  Right
  panel: velocity gradients ($dw$/$du$) as a function of $u$ for the
  same solutions presented in the left panel.  Note the change in the
  characteristic shape of the profile that occurs for $k$ values
  between 2.0 and 3.0.
\label{fig:k}}
\end{figure}

\begin{figure}
\plotone{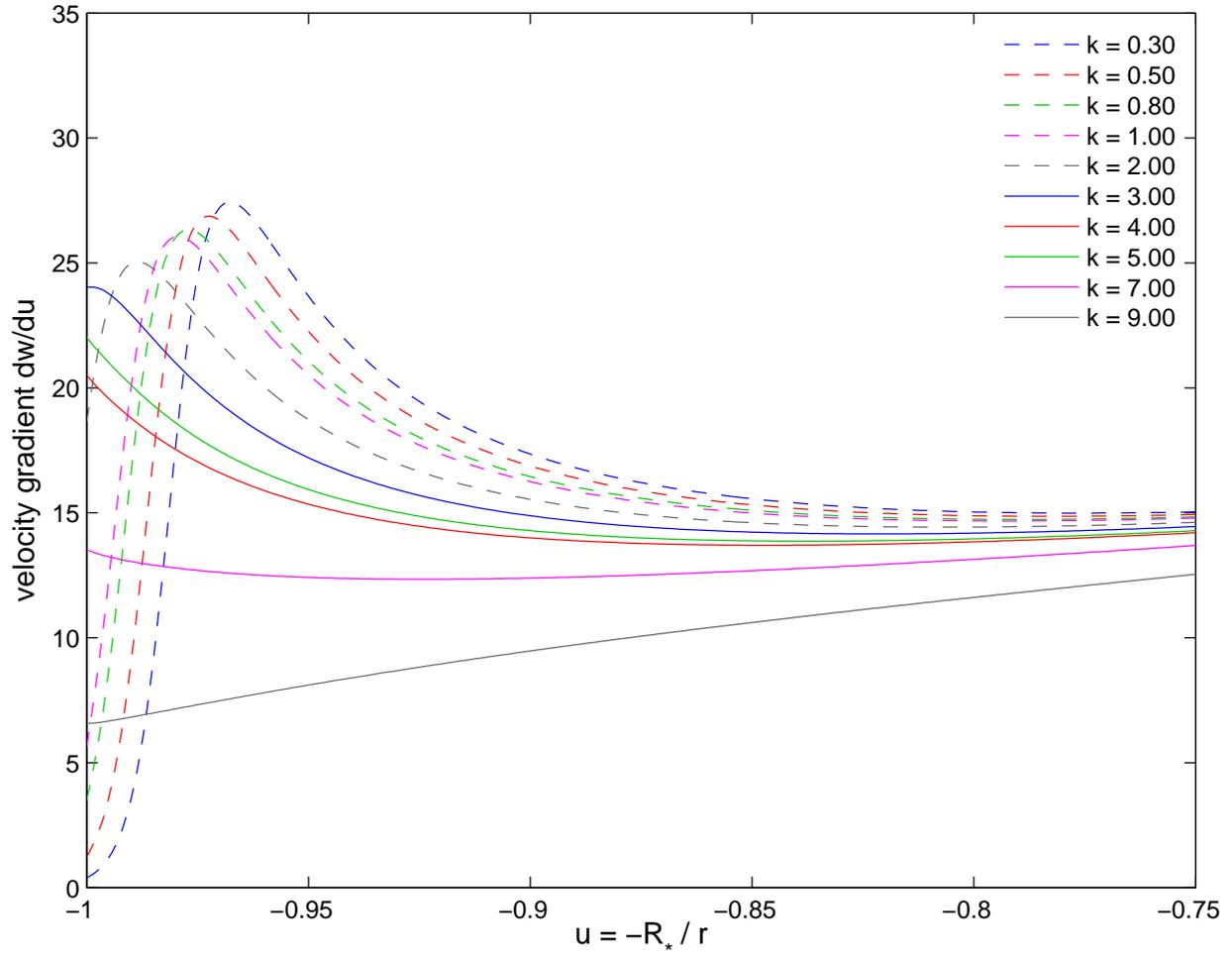}
\caption{Magnified view of the behavior near the stellar surface of
  the velocity gradients shown in the right panel of
  Figure~\ref{fig:k}. \label{fig:k2}}
\end{figure}

Figure~\ref{fig:k} is similar to Figure~\ref{fig:alpha}, depicting the
velocity structure and its gradient as a function of $u$ for the $k$
values listed in Table~\ref{table:B1b}.  As shown in the left panel of
Figure~\ref{fig:k}, increasing the $k$ value results primarily in higher
initial velocities at the stellar surface, but does not affect the
terminal velocity or the characteristic behavior of the velocity
profile overall.  (However, we note that for $k \ga 5.0$, the initial
velocity is probably too high to be physical.) In contrast, the
velocity \textit{gradient} (shown in the right panel of
Figure~\ref{fig:k}) completely loses its characteristic hump near the
stellar surface somewhere between $2 < k < 3$.  Figure~\ref{fig:k2}
shows a magnified view of the velocity gradient near the stellar
surface so that the transition to the new kind of behavior may be
more clearly seen.  The solutions corresponding to high $k$ values
(i.e., $> 2$ for this choice of $\alpha$ and $\delta$) and different
velocity gradient structures (i.e., that do not display the
characteristic hump) were previously unknown.  To distinguish these
high $k$, high $\Omega$ solutions from the other solutions, we refer
to them as the $k\Omega$ solutions.

\begin{figure}
\plottwo{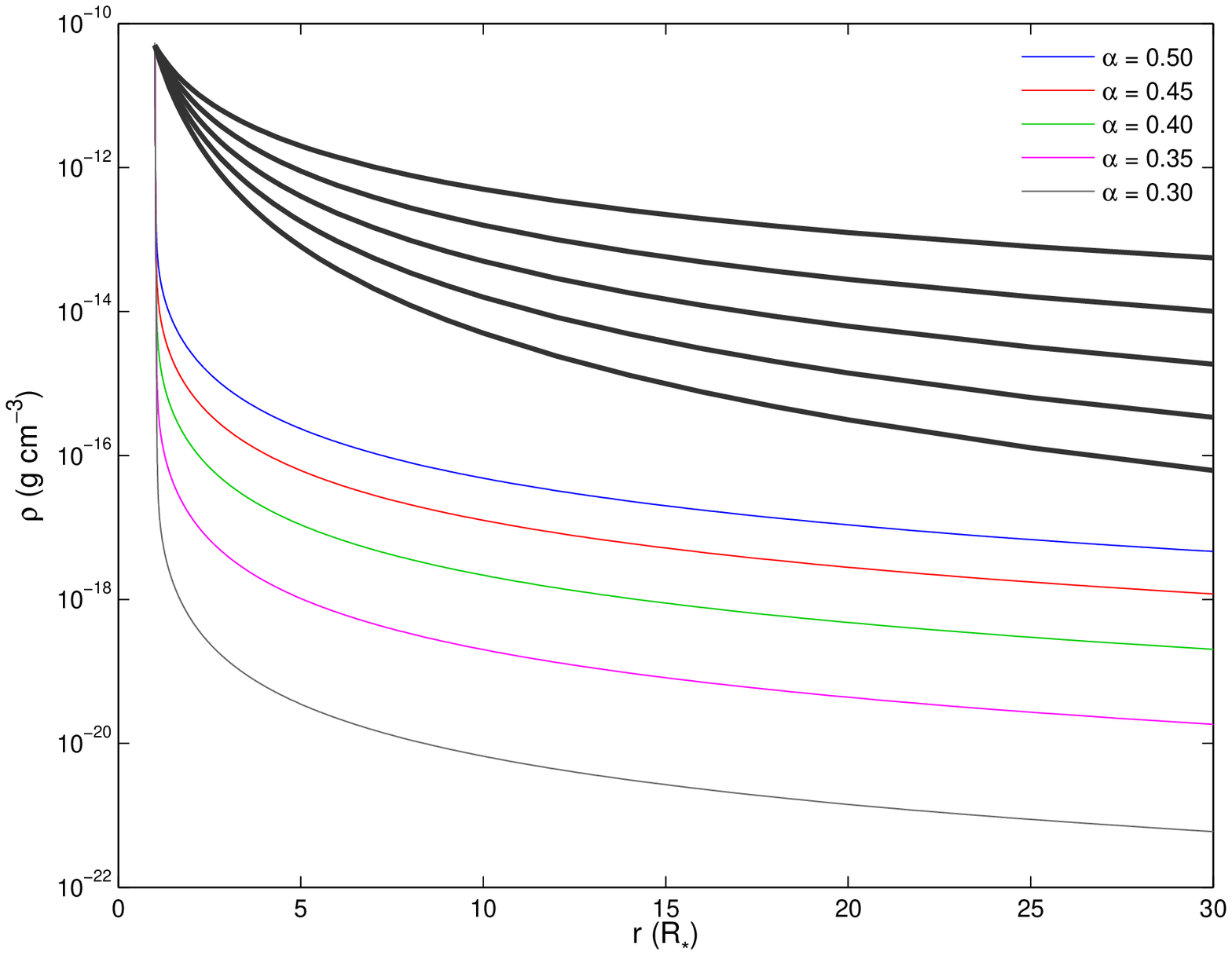}{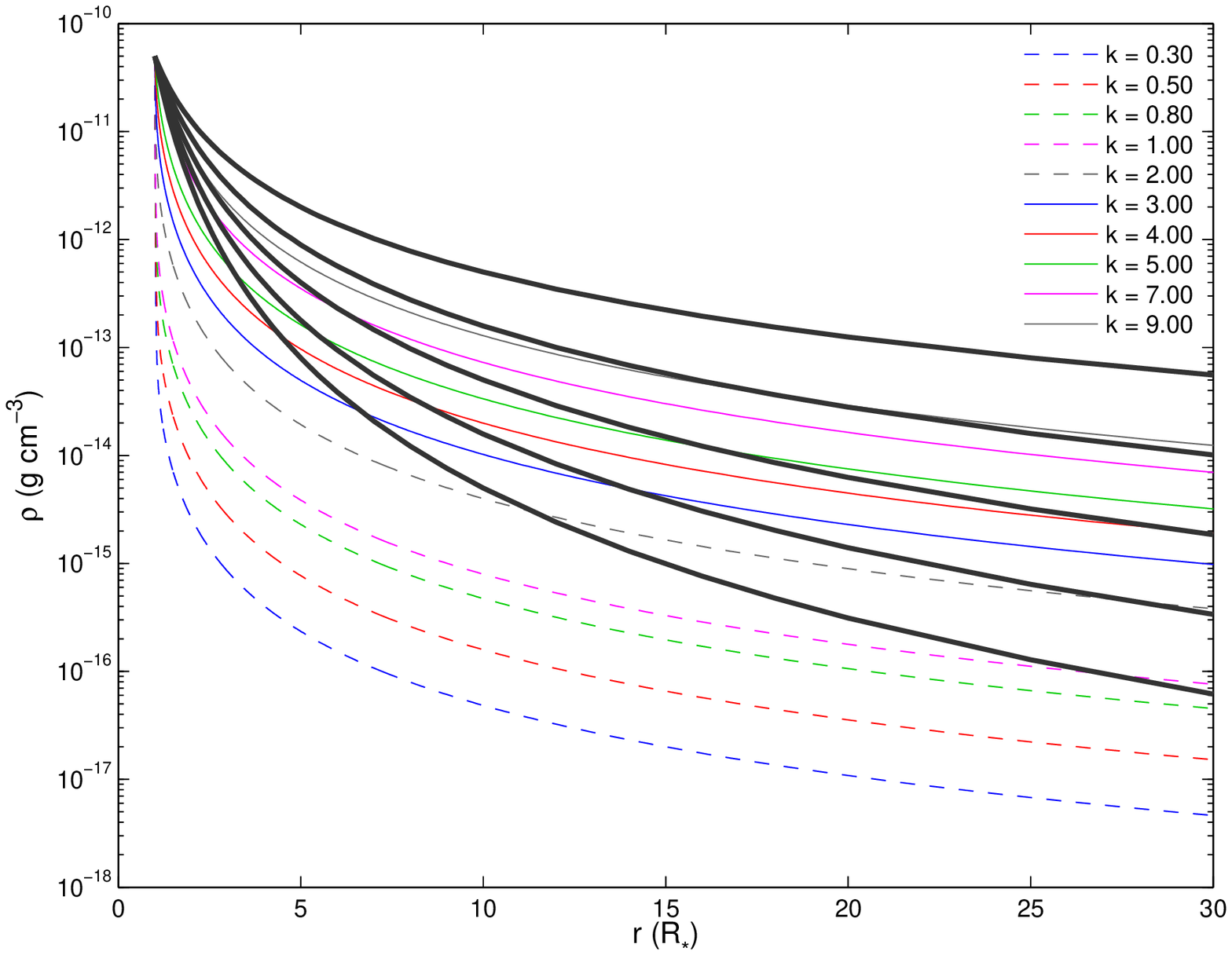}
\caption{Comparison of equatorial density as a function of stellar
  radius for various \textsc{Bedisk} and \textsc{Hydwind} models. Left
  panel: $k = 0.3$ and $\delta = 0.07$ are held fixed while $\alpha$
  varies from 0.5 to 0.3, in increments of 0.05, from top to bottom,
  shown in thin lines.  Above this, ad hoc density distributions that
  follow a simple power-law drop-off, from $n = 2.0$ to 4.0 from top
  to bottom, in increments of 0.5, are shown in thick lines.  In both
  cases, the density at the stellar surface ($\rho_* = \rho_0$) is set
  to $5 \times 10^{-11}$ g\,cm$^{-3}$. Right panel: $\alpha = 0.5$ and
  $\delta = 0.07$ are held fixed while $k$ takes the values 0.3, 0.5,
  0.8, 1.0, 2.0, 3.0, 4.0, 5.0, 7.0, and 9.0, from bottom to top,
  shown in thin lines. Again, the same power-law governed equatorial
  densities shown in the left panel are shown in thick
  lines. \label{fig:density}}
\end{figure}

Tables~\ref{table:B1a} and ~\ref{table:B1b} give some indication (when
starting from the line force parameters of \citealt{abb82}) of which
combinations of parameters yield mass-loss rates and terminal
velocities similar to those predicted for Be stars. This information
can be combined to generate the expected equatorial density structure
that would result from the wind, which is something that can be
directly compared to the ad hoc density structures typically assumed
for Be stars.  In Figure~\ref{fig:density}, the equatorial densities
computed from the line-driven wind models are compared with ad hoc
models of Be star disks.  In both panels, typical ad hoc equatorial
density profiles, governed by $n = 2.0, 2.5, 3.0, 3.5,$ and $4.0$ (from
top to bottom) are shown in thick lines, while the equatorial density
profiles computed from the hydrodynamic equations (with different
values of the line force parameters) are shown in thin lines.

The left panel of Figure~\ref{fig:density} depicts the solutions of
constant $k = 0.3$ and $\delta = 0.07$, with $\alpha$ varying from 0.5
to 0.3 (in increments of 0.05) from top to bottom.  Clearly, even the
solution for the highest value of $\alpha$ has density values that are
significantly lower than the density values that are typically
assumed, and that have been shown to reproduce the observed H$\alpha$
emission signature.  While lowering the value of $\alpha$ results in
terminal velocities that better match those of Be stars, it obviously
increases the discrepancy between the two density structures.
Synthetic line profiles computed for these solutions only produce the
stellar absorption profile, because, as expected, these densities are
too low to produce any significant amount of emission.

\begin{figure}
\plotone{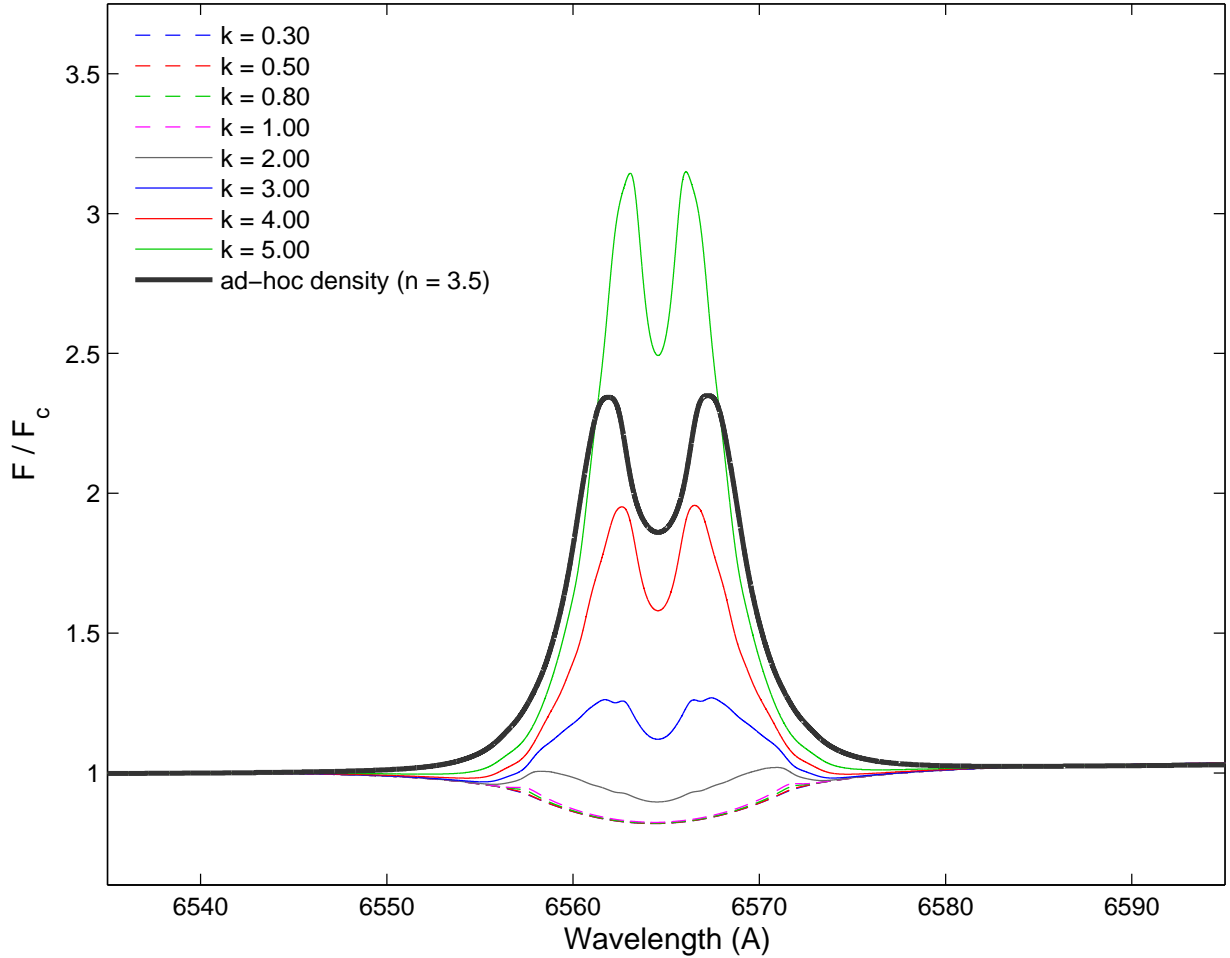}
\caption{Synthetic H$\alpha$ profiles computed from the equatorial
  density structures obtained from the solution of the hydrodynamic
  equations and used as input into the radiative transfer code
  \textsc{Bedisk}, shown in thin lines, for $k$ values of 0.3, 0.5,
  0.8, 1.0, 2.0, 3.0, 4.0, and 5.0. For comparison, an emission profile
  computed from the ad hoc density structure $\rho = 5\times10^{-11}$
  g\,cm$^{-3} (r/R_*)^{-3.5}$ is also shown. An inclination of $i =
  35\degr$ was assumed for all profiles. Clearly, $k \le 1$ produces
  only an absorption profile, but emission begins for $k > 1.0$ and
  becomes progressively stronger for higher $k$
  values. \label{fig:profiles}}
\end{figure}

The right panel of Figure~\ref{fig:density} depicts the solutions of
constant $\alpha = 0.5$ and $\delta = 0.07$, with $k$ increasing from
0.3, to 0.5, 0.8, 1.0, 2.0, 3.0, 4.0, 5.0, 7.0, and 9.0, from bottom
to top.  What is seen in this plot is that for $k$ values $\sim2$ and
greater, the equatorial densities computed by the two different
methods become comparable.  Using these equatorial densities as input
into \textsc{Bedisk}, significant emission was found for $k \ge 2$.
This is depicted in Figure~\ref{fig:profiles}, which shows the
emission profiles for $k$ = 0.3, 0.5, 0.8, 1.0, 2.0, 3.0, 4.0, and 5.0
in thin lines.  The lowest $k$ values ($k \le 1$) produce only the
stellar absorption profile, with the emission first becoming
appreciable at $k = 2$ and increasing in strength with increasing $k$.
For comparison, a synthetic line profile created from an ad hoc
equatorial density structure with $\rho_0 = 5 \times 10^{-11}$
g\,cm$^{-3}$ and $n = 3.5$ (the value of $n$ most commonly found for
Be stars) is shown in the thick line.

\begin{figure}
\plotone{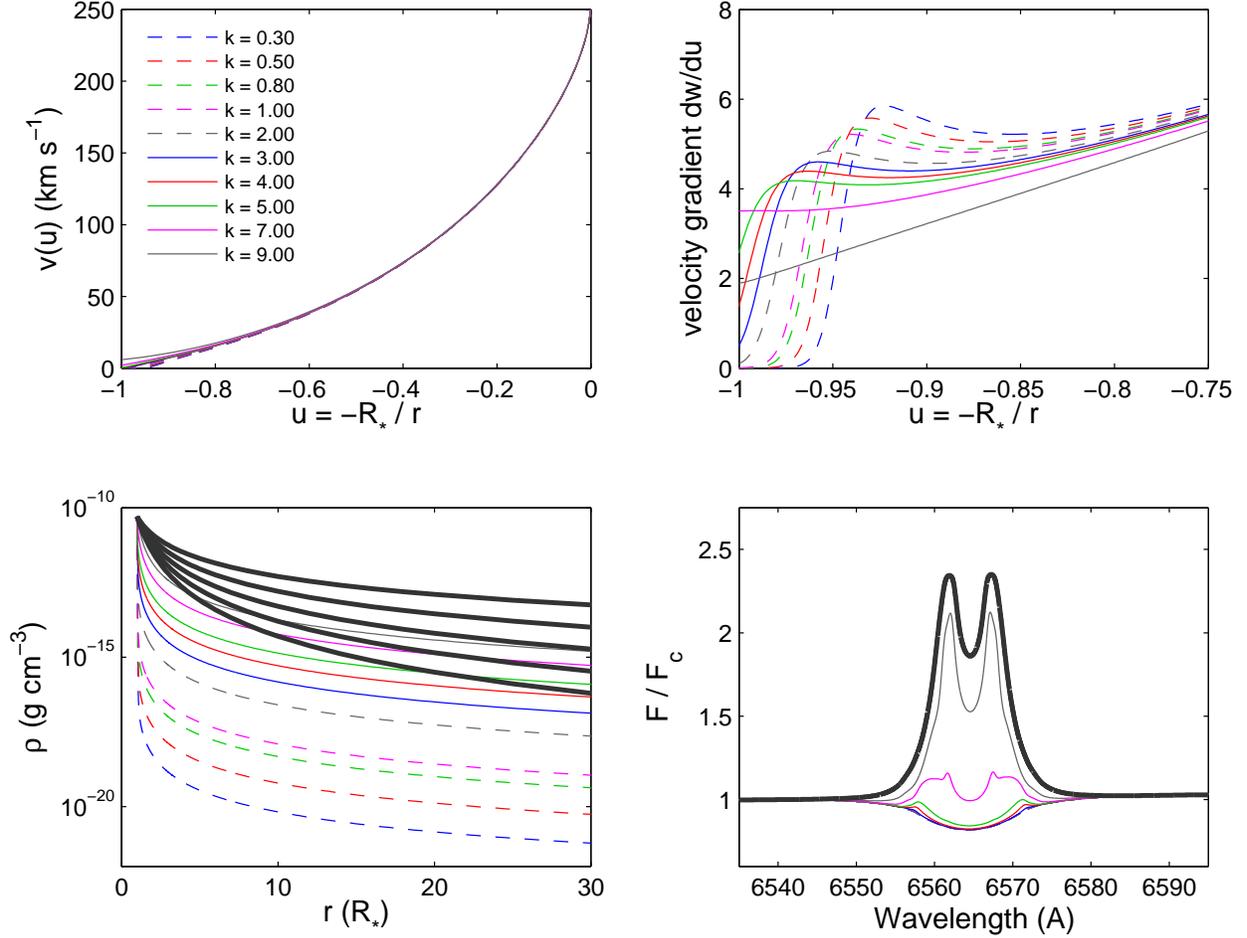}
\caption{Results for simulations performed using a fixed $\alpha$
  value of 0.3, $\delta$ = 0.07, and various $k$ values as given in
  the legend in the top left panel.  The legend applies to all plots
  in the figure. In the lower left panel, the thick, dark gray lines
  correspond to the same ad hoc density structures shown in
  Figure~\ref{fig:density}, and in the lower right panel, the thick,
  dark line corresponds to the same emission profile from an ad hoc
  density structure that is shown in
  Figure~\ref{fig:profiles}. \label{fig:lowalpha}}
\end{figure}

From the above investigation, we deduce that the lowest value of
$\alpha$ used in this study ($\alpha = 0.3$) may better describe Be
star disks because of its low outflow velocity; thus, we have further
explored this parameter space.  In Table~\ref{table:B1c}, we show the
mass-loss rates and terminal velocities for solutions with $\alpha =
0.3$ and $\delta = 0.07$ held fixed while $k$ is again allowed to
vary.  In Figure~\ref{fig:lowalpha}, we show the characteristics of
these solutions. The velocities (shown in the top left panel of
Figure~\ref{fig:lowalpha}) of these low-$\alpha$ solutions are
remarkably constant over the whole range of $k$ values that we
employed. We note that, even at high $k$ values, the solutions do not
obtain high velocities at the stellar surface (i.e., they do not
  become unphysical), which represents a marked contrast to the
solutions obtained with $\alpha$ fixed at 0.5.  In the top right panel
of this same figure, we show the velocity gradients of the $\alpha$ =
0.3 solutions in the region close to the stellar surface.  Solutions
of the new $k\Omega$-type are found for $k = 7.0$ and $k = 9.0$. In
the bottom left panel, the densities of the $\alpha = 0.3$ solutions
are compared with typical ad hoc equatorial densities corresponding to
$n = 2.0$ (topmost dark gray line) to $n = 4.0$ (bottom-most dark gray
line) and changing by increments of 0.5 (similarly to
Figure~\ref{fig:density}.)  Finally, in the bottom right panel, the
emission expected from these equatorial densities is shown.  In the
$\alpha = 0.3$ case, we see significant emission only occurring from
$k = 7.0$ and higher.  This corresponds to the same value of $k$ that
marks the emergence of the $k\Omega$ solutions.
 
\section{Summary and Discussion}

Previous attempts to model Be stars in the context of radiatively
driven winds added a centrifugal force term to the momentum equation,
to account for the effect of rapid rotation, but still drew solutions
from the family of solutions originally obtained by \citet{cas75} in
their pioneering work.  Indeed, \citet{cas75}, who had neglected
rotation entirely in their model, had shown the existence of only one
family of physical solutions.  \citet{cur07} confirmed that the
standard m-CAK singular point is the only one that satisfies the
boundary condition at the stellar surface when rotation is
neglected. All such solutions, however, have terminal velocities that
are too high to be compatible with those found in Be star disks.

When \citet{cur04} performed a reanalysis of the hydrodynamic
equations with a momentum equation that included the centrifugal force
term, he found a new family of solutions (the $\Omega$-slow solutions)
with much lower terminal velocities and higher mass-loss rates.  He
showed that while these new solutions are in fact present for all
rotation rates, the traditional CAK solutions actually cease to exist
for $v_{\rm{rot}}/v_{\rm{crit}} \approx 0.75$, and therefore these new
solutions are the only ones that can exist for very high rotation
rates.  The existence of a slowly outflowing, dense wind for
$v_{\rm{rot}} \ge 0.75v_{\rm{crit}}$ seemed immediately appropriate
for a Be star, and we therefore felt it merited further investigation.

We started our investigation from the line-force parameters for a B1V
star ($T_{\rm{eff}} = 25\,000$~K) as given by \citet{abb82}: $k$ =
0.3, $\alpha$ = 0.5, and $\delta$ = 0.07.  These parameters yield a
terminal velocity that is slightly too high for Be stars (by about a
factor of two), and a mass-loss rate that is insufficient to build up
an equatorial density structure that produces significant emission.
We recall, however, that the parameters derived by Abbott are done so
under the assumptions of the original CAK theory, and may no longer be
appropriate for the new family of $\Omega$-slow solutions.  Thus, we
systemically varied $\alpha$ and $k$ to see what values of these two
parameters produce terminal velocities and mass-loss rates that agree
with the values found for Be stars.  We have shown that if $k$ and
$\delta$ are fixed at 0.3 and 0.07, respectively, then $\alpha$ = 0.3
produces $v_{\infty} = 256$~km\,s$^{-1}$.  This is in good agreement
with \citet{poe78}, who suggested that the outflow velocity of a Be
star disk was of the order of $\sim$200~km\,s$^{-1}$.  Alternately, if
$\alpha$ and $\delta$ are held fixed at 0.5 and 0.07, respectively,
$k$ values of 2.0 and greater produce equatorial densities comparable
to the ones assumed in the ad hoc scenario, and also produce
appreciable emission if those equatorial density structures are used
as input in \textsc{Bedisk}.  

Finally, because solutions with $\alpha = 0.3$ had the best agreement
with the terminal velocity predicted for Be stars, we further explored
this parameter space.  By allowing $k$ to vary over the same range as
in our previous experiment, we found that all solutions with $\alpha =
0.3$ retained low velocities near the stellar surface (meaning that
high $k$ values did not lead to unphysical solutions in this case).
By examining the velocity gradient of these solutions, it was found
that the $k\Omega$ solutions emerged around $k \gtrsim 7.0$, which is
considerably higher than when $\alpha = 0.5$ ($k\Omega$ solutions
emerge at $k \gtrsim 2$).  Furthermore, only solutions with $k = 7.0$
or higher produced significant emission in our simulations. However,
as outlined in the Future Work section (see below), we feel there are
additional alterations and refinements that we could apply to our
model that may allow us to achieve larger equatorial densities from
smaller $k$ values.

One aspect of our calculation that merits additional discussion is
that, because \textsc{Bedisk} considers the disk to be in Keplerian
rotation, and considers this to be the \textit{only} source of profile
broadening, outflow in the disk is neglected.  However, the bulk of
the H$\alpha$ emission has been shown to arise from the region
$\sim5-15 R_*$ from the stellar surface, and as shown in Table 3 of
\citet{poe78}, the outflow velocities at $r/R_*$ = 6.0 and $r/R_*$ =
18.0 are predicted to be 25~km\,s$^{-1}$ and 125~km\,s$^{-1}$,
respectively.  Therefore, especially for the innermost region of
H$\alpha$ emission, where densities are highest (and thus the majority
of the emission is produced), the outflow velocity is negligible
compared with the rotational velocity.

In performing the systematic investigation of the line-force
parameters, we discovered a new behavior in the velocity gradient for
solutions with high $k$ values, which we call the $k\Omega$ solutions.
This new behavior was observed to emerge between $2 \le k \le 3$,
when $\alpha$ and $\delta$ are 0.5 and 0.07, respectively. While $k$
has traditionally been found to have a value of $\sim0.5$ for a star
with the effective temperature we have adopted in our model, we recall
that all of the previous analyses have been performed in the regime of
the fast winds that pass through the original m-CAK critical point.
In a private communication with Joachim Puls (2014), he stated that,
generally, $k \sim1/\alpha$; since we employ values of $\alpha = 0.5$
to $0.3$, $k = 2$ or $k \gtrsim 3$ are predicted from this
relation. When $\alpha$ = 0.5 and $k \ge 5$, however, we find
that the solutions become unphysical due to the high initial
velocities that they exhibit.

\section{Future Work}

We have begun our investigation into the possible role of the
$\Omega$-slow solutions in the formation of Be star disks by first
examining the \textit{equatorial} density structure that arises from
such winds, i.e., we have examined the 1D solutions. We have compared
these equatorial density structures to the ad hoc scenario in which a
given initial density is assumed to fall off according to a single
power law with increasing radial distance from the star.  In order to
estimate the magnitude of the \ha emission that might arise from
$\Omega$-slow solutions, various equatorial density structures
computed from these solutions were used as input into the code
\textsc{Bedisk}.  We have shown that significant H$\alpha$ emission
can be produced by some combinations of the line-force parameters.

In a future paper, we plan to do a comparison of the full 2D structure
computed from the hydrodynamic equations versus the 2D structure
computed by \textsc{Bedisk}.  If the vertical structure computed by
\textsc{Hydwind} has more material in the regions close to the
equatorial plane (i.e., slightly above and below it), then it may be
possible to produce significant emission in models with lower $k$
values than the ones used here.

An additional important consideration may be the inclusion of the
oblate finite disk correction factor.  For simplicity, we have started
from the usual finite disk correction factor, which approximates the
star as a uniformly bright sphere.  Realistically, however, when a
star is rotating close to its critical velocity (such as in the case
of Be stars), the star becomes an oblate spheroid, with a cooler
equatorial region and hotter poles.  As shown in \citet{ara11}, the
use of an oblate finite disk correction factor results in higher mass
loss rates in the equatorial plane.  Thus, this may represent yet
another way to reproduce emission profiles with smaller $k$ values
than we have employed in this work.

\acknowledgments

We thank the anonymous referee, whose comments helped to strengthen
and improve this paper.  We also thank Joachim Puls and Lydia Cidale
for their helpful discussions on this project.  This work has made use
of NASA's Astrophysics Data System.  M.C. acknowledges, with thanks,
the support of FONDECYT project 1130173 and Centro de Astrof\'{i}sica
de Valpara\'{i}so.  This research was supported in part by NSERC, the
Natural Sciences and Engineering Research Council of Canada.

\clearpage

\begin{deluxetable}{cccccccc} 
\tablecolumns{3} 
\tablewidth{0pc} 
\tablecaption{B1V Wind Models with $k = 0.3$, $\delta = 0.07$, and Varying $\alpha$ Values.  \label{table:B1a}} 
\tablehead{ 
\colhead{$\alpha$} & \colhead{$\dot{M}$} & \colhead{$v_{\infty}$}  \\
\colhead{} & \colhead{(10$^{-6} M_{\sun}\,yr^{-1}$)} & \colhead{(km\,s$^{-1}$)}
} 
\startdata 
0.50    &  4.273$\times 10^{-3}$ & 430    \\ 
0.45    &  9.632$\times 10^{-4}$ & 381    \\ 
0.40    &  1.428$\times 10^{-4}$ & 337    \\ 
0.35    &  1.116$\times 10^{-5}$ & 295    \\ 
0.30    &  3.038$\times 10^{-7}$ & 256    \\   
\enddata 
\tablecomments{All models assume an initial density $\rho_* = 5 \times 10^{-11}$ g\,cm$^{-3}$.}
\end{deluxetable}

\begin{deluxetable}{cccccccc} 
\tablecolumns{3} 
\tablewidth{0pc} 
\tablecaption{B1V Wind Models with $\alpha = 0.5$, $\delta = 0.07$, and Varying $k$ Values.  \label{table:B1b}} 
\tablehead{ 
\colhead{$k$} & \colhead{$\dot{M}$} & \colhead{$v_{\infty}$} \\
\colhead{} & \colhead{(10$^{-6} M_{\sun}\,yr^{-1}$)} &\colhead{(km\,s$^{-1}$)}
} 
\startdata 
0.30    &  4.273$\times 10^{-3}$ & 430   \\ 
0.50    &  1.402$\times 10^{-2}$ & 430   \\ 
0.80    &  4.182$\times 10^{-2}$ & 430   \\ 
1.00    &  7.026$\times 10^{-2}$ & 430   \\ 
2.00    &  3.522$\times 10^{-1}$ & 430   \\
3.00    &  9.042$\times 10^{-1}$ & 430   \\ 
4.00    &  1.765$\times 10^{0}$  & 430   \\ 
5.00    &  2.966$\times 10^{0}$  & 430   \\ 
7.00    &  6.482$\times 10^{0}$  & 432   \\ 
9.00    &  1.161$\times 10^{1}$  & 435   \\ 
\enddata 
\tablecomments{All models assume an initial density $\rho_* = 5 \times 10^{-11}$ g\,cm$^{-3}$.}
\end{deluxetable} 
\clearpage

\begin{deluxetable}{cccccccc} 
\tablecolumns{3} 
\tablewidth{0pc} 
\tablecaption{B1V Wind Models with $\alpha = 0.3$, $\delta = 0.07$, and Varying $k$ Values. \label{table:B1c}} 
\tablehead{ 
\colhead{$k$} & \colhead{$\dot{M}$} & \colhead{$v_{\infty}$} \\
\colhead{} & \colhead{(10$^{-6} M_{\sun}\,yr^{-1}$)} &\colhead{(km\,s$^{-1}$)}
} 
\startdata 
0.30    &  3.038$\times 10^{-7}$ & 256   \\ 
0.50    &  2.800$\times 10^{-6}$ & 256   \\ 
0.80    &  2.160$\times 10^{-5}$ & 256   \\ 
1.00    &  5.700$\times 10^{-5}$ & 256   \\ 
2.00    &  1.161$\times 10^{-3}$ & 256   \\
3.00    &  6.765$\times 10^{-3}$ & 256   \\ 
4.00    &  2.363$\times 10^{-2}$  & 256   \\ 
5.00    &  6.235$\times 10^{-2}$  & 256   \\ 
7.00    &  2.692$\times 10^{-1}$  & 256   \\ 
9.00    &  8.028$\times 10^{-1}$  & 256   \\ 
\enddata 
\tablecomments{All models assume an initial density $\rho_* = 5 \times 10^{-11}$ g\,cm$^{-3}$.}
\end{deluxetable} 
\clearpage

\end{document}